\documentclass[conference]{IEEEtran}
\IEEEoverridecommandlockouts

\usepackage{cite}
\usepackage{amsmath,amssymb,amsfonts}
\usepackage{algorithmic}
\usepackage{graphicx}
\usepackage{textcomp}
\usepackage{xcolor}

\usepackage{xurl}
\usepackage{ulem}
\usepackage{caption}
\usepackage{hyperref}
\usepackage{multirow}
\usepackage{tabularx}
\usepackage{longtable}
\usepackage{algorithm}
\usepackage{threeparttable}

\begin{document}

\title{A Cascade Approach for APT Campaign Attribution in System Event Logs: Technique Hunting and Subgraph Matching}

\author{Yi-Ting Huang, Ying-Ren Guo, Guo-Wei Wong, and Meng Chang Chen}

\maketitle

\begin{abstract}
As Advanced Persistent Threats (APTs) grow increasingly sophisticated, the demand for effective detection methods has intensified. This study addresses the challenge of identifying APT campaign attacks through system event logs. A cascading approach, name \textit{SFM}, combines Technique hunting and APT campaign attribution. Our approach assumes that real-world system event logs contain a vast majority of normal events interspersed with few suspiciously malicious ones and that these logs are annotated with Techniques of MITRE ATT\&CK framework for attack pattern recognition. Then, we attribute APT campaign attacks by aligning detected Techniques with known attack sequences to determine the most likely APT campaign. Evaluations on five real-world APT campaigns indicate that the proposed approach demonstrates reliable performance.
\end{abstract}

\begin{IEEEkeywords}
Intrusion Detection, Technique Hunting, APT Campaign Detection 
\end{IEEEkeywords}

\section{Introduction}
The rise of sophisticated Advanced Persistent Threats (APTs) has posed significant challenges to the cybersecurity community. High-profile attacks like BlackEnergy~\cite{blackenergy2015ukraine} and the SolarWinds Compromise~\cite{fireeye2020solarwind} have demonstrated the evolving nature of these attacks. While there has been progress in malware analysis~\cite{MAMBA,APILI}, the detection and mitigation of APTs remain urgent priorities.
APTs are typically orchestrated by advanced threat actors with economic or political motivations. APT campaigns differ from traditional malware or botnet attacks in that they are multistage operations that often begin with gaining a foothold in a target environment, followed by prolonged periods of undetected activity, data exfiltration, and system compromise. Given the complexity and stealth of APT operations, developing detection mechanisms that can identify known campaigns has become a critical priority in the cybersecurity field, ensuring effective responses to these advanced threats.

Recent advances in host-based intrusion detection systems (HIDS) have been thoroughly reviewed in a recent survey~\cite{michael2022provenance}. To address the persistent and multifaceted nature of Advanced Persistent Threats (APTs), systems like Holmes~\cite{milajerdi2019holmes} and MORSE~\cite{hossain2020combating} have shown that combining coarse-grained analysis (which classifies events as benign or malicious) with fine-grained analysis (which maps events to Tactics, Techniques, and Procedures, TTPs) can significantly enhance threat detection capabilities.
In addition, approaches like RepSheet~\cite{hassan2020tactical} and KRYSTAL~\cite{kurniawan2022krystal} focus on detecting known attack descriptors to construct contextual attack scenarios, further improving understanding of intrusion activity. However, these methods typically require manual input to define mapping rules for recognizing attack patterns, which limits their scalability and automation potential.

Forensic analysis of security incidents, whether to attribute attacks to specific threat actors or align them with known campaigns based on observable artifacts, remains a labor-intensive process. Once an attack is detected, human analysts must sift through vast amounts of event logs to trace the compromise, prioritize alerts, and attribute the attack to a particular actor or group by comparing it to previous campaigns.
Although few studies have explored cyber threat attribution based on observable attack stages~\cite{sachidananda2023apter}, attacker profiling~\cite{ren2022cskg4apt}, or artifact analysis~\cite{sahoo2022cyber} and the attribution process often points to specific actors, recognizing intrusion activities as part of known APT campaigns is equally important for improving system defenses and accelerating incident response.

To address the challenge, we propose a StraightForward Method (SFM) for audit log analysis and APT campaign detection.
We construct a machine learning-based method to discover malicious behaviors and ultimately identify potential APT threats. 
Specifically, the tasks of this study are:
1) design a neural network detection model to discover malicious behaviors and 2) identify the most likely APT campaign by matching the discovered behaviors with known APT campaigns.

\section{Background and Motivation}\label{sec:background}
\subsection{Motivating Example}
\begin{figure*}[!h]
    \centering
    \includegraphics[width=0.8\textwidth]{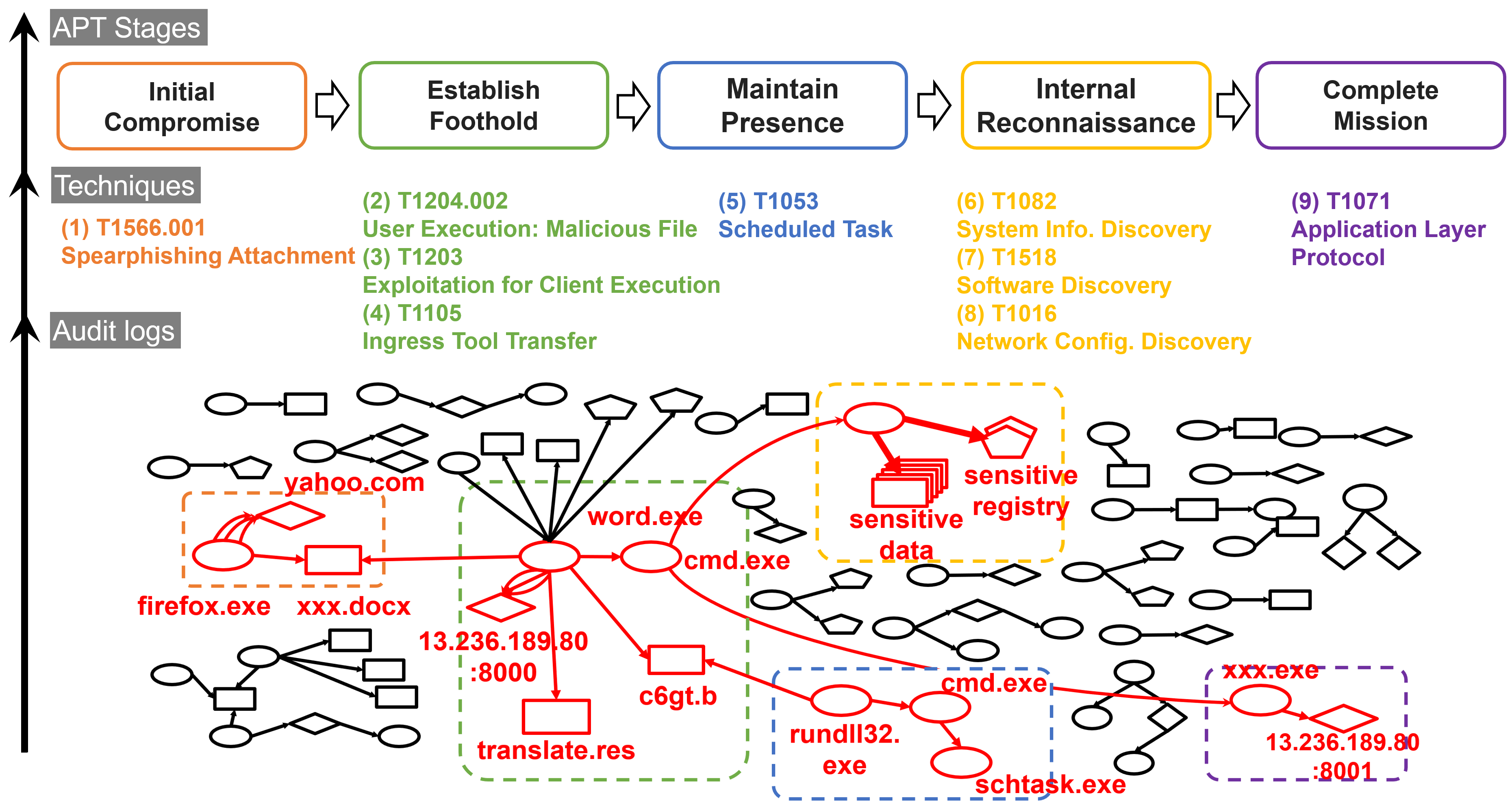}
    \caption{SharpPanda: motivating example}
    \label{fig:sharppanda}
\end{figure*}
Figure~\ref{fig:sharppanda} illustrates the provenance graph of a sophisticated and persistent attack SharpPanda~\cite{SharpPanda} that uses spear-phishing tactics to launch cyberattacks on government officials.
The attack begins with the victim receiving a phishing email containing a malicious attachment, ``
...trackchanged.docx.'' Upon execution of the file, it establishes a connection to a remote server controlled by the attacker, from which additional payloads are downloaded. These payloads exploit several known vulnerabilities, including CVE-2018-0802, CVE-2018-0798, and CVE-2017-11882, leading to the deployment of a DLL file on the victim's system. The attacker then configures a scheduled task to ensure persistence and executes a series of commands to collect system and network information. Finally, the gathered data is exfiltrated to the attacker's server.

To illustrate our approach to detect APT campaigns, this example highlights several key challenges:
\begin{itemize}
    \item Imbalance data handling,
    \item Technique identification,
    \item APT Campaign attribution.
\end{itemize}

\subsection{Event Log}
\begin{figure}[!htb]
    \centering
    \includegraphics[width=0.48\textwidth]{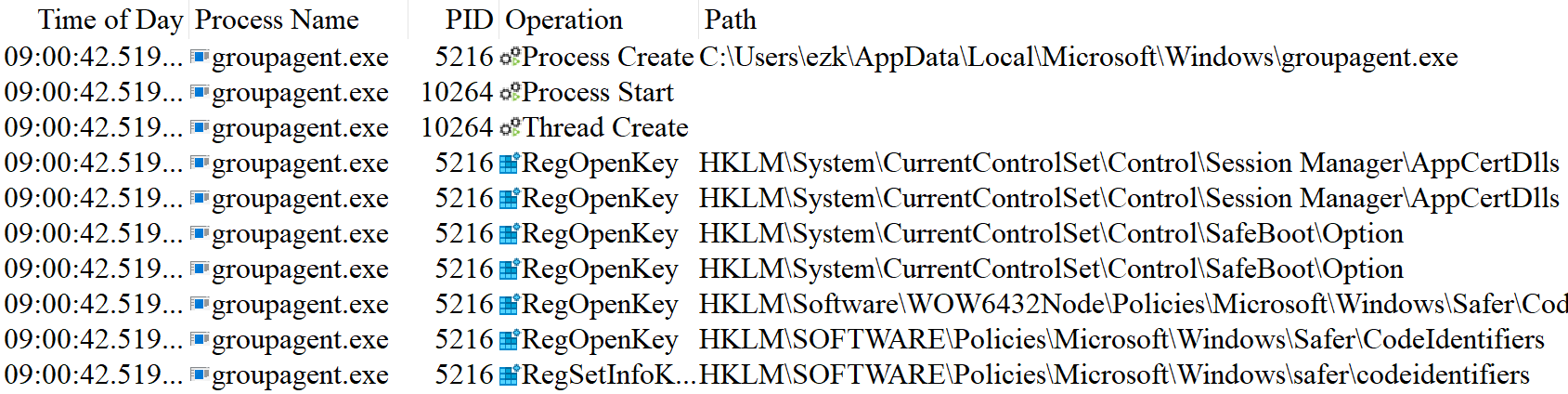}
    \caption{Example of audit log events captured by \textit{Procmon}.}
    \label{fig:procmon}
\end{figure}

An event log is a system-generated record that documents activities related to system operations. Logs collected via mechanisms like Windows Event Tracing (ETW), Linux Audit, or tools such as ProcMon and Sysmon provide critical insights for security monitoring and incident response. An entry in the audit log, called an event, contains detailed information such as timestamps, event sources, event types, process IDs, and other information that enables security analysts to capture and analyze system operations in real time, as illustrated in Figure~\ref{fig:procmon}. 
In this example, the process \textit{groupagent.exe} (PID 5216) is active on the system, creating a child process (PID 10264) and accessing multiple system registry files.

\subsection{MITRE ATT\&CK Framework}
MITRE ATT\&CK is a comprehensive knowledge base of tactics, techniques, and procedures (TTPs) used by adversaries in cyberattacks. The framework covers various attack methods employed by attackers from initial system intrusion to achieving their objectives. By analyzing system logs and mapping anomalies to attack techniques, we can link them to adversary groups and develop automated detection models for specific APT attack strategies.

\subsection{APT lifecycle}
The APT (Advanced Persistent Threat) attack lifecycle involves several stages, from initial compromise to achieving objectives. It provides a structured way to analyze APTs and integrate frameworks like MITRE ATT\&CK for better threat detection. In this study, we reference the Mandiant attack lifecycle, which includes seven flexible stages: Initial Compromise, Establishing Foothold, Reconnaissance, Privilege Escalation, Lateral Movement, Persistence, and Mission Completion.

\section{A Straight Forward Method for APT Campaign Detection}\label{sec:SFM}
\begin{figure}[!htb]
    \centering
    \includegraphics[width=0.48\textwidth]{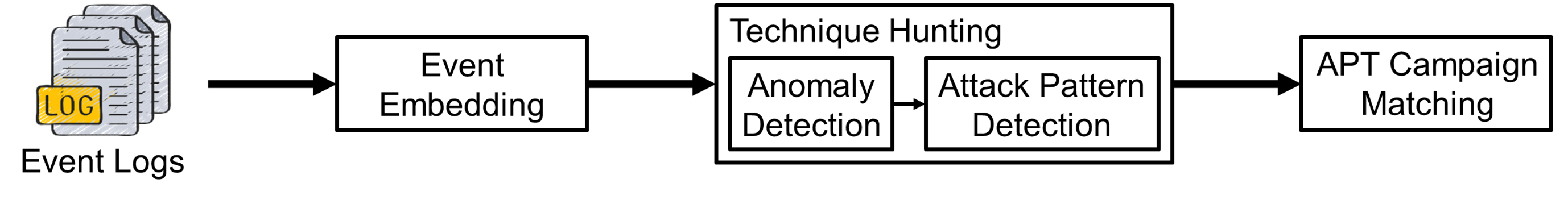}
    \caption{SFM workflow}
    \label{fig:workflow}
\end{figure}

Currently, there are few, if any,  APT campaign detection methods that effectively identify embedded Techniques within audit logs and subsequently use these discovered Techniques to detect APT campaigns. To address this challenge, we propose a straightforward method (SFM) that employs deep learning models to automatically identify and analyze attack techniques from audit logs, which then enables inference of potential APT campaigns for a more comprehensive detection result. Figure~\ref{fig:workflow} illustrates the workflow of the SFM model.

SFM consists of three main stages: event embedding, Technique hunting, and APT campaign matching. First, audit events are processed to construct a data provenance, ensuring that causal relationships are maintained in the analysis. Next, a neural network model is introduced for Technique hunting, designed specifically to identify attack patterns (Section~\ref{sec: TTP hunting}). Finally, the system evaluates the likelihood between the identified attack behaviors and known APT campaigns to assess if an attack is unfolding (Section~\ref{sec: campaign match}). Although this initial approach shows promise, further development is expected to enhance the performance of SFM.

\begin{figure}[htb!]
    \centering
    \includegraphics[width=0.48\textwidth]{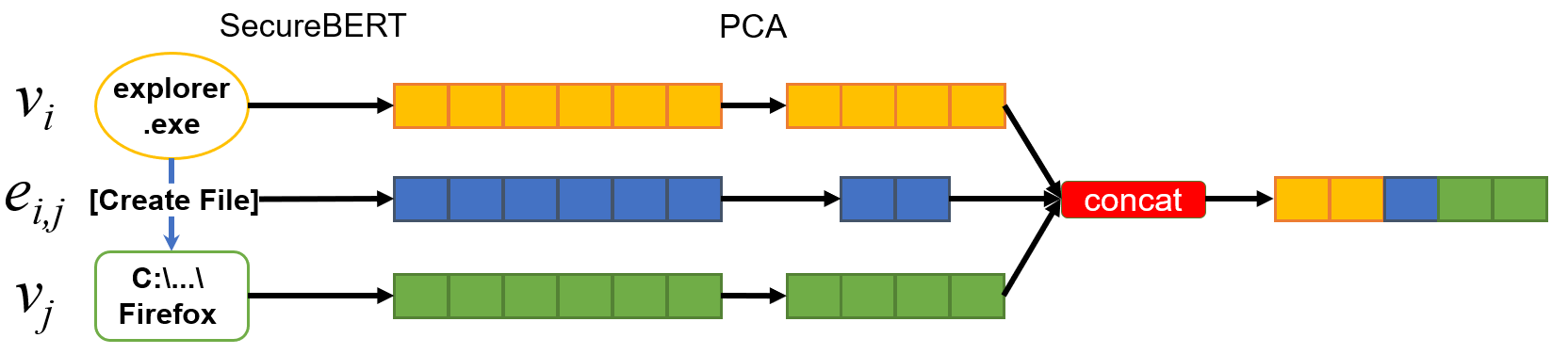}
    \caption{Embedding of an audit log event.}
    \label{fig:triplet_embedding}
\end{figure}

\subsection{Event Embedding}\label{sec: embedding}
First, event logs are collected from the Process Monitor (ProcMon), which records detailed system activities such as process creation and file access. These logs provide critical information for analyzing system behavior. For example, the fifth event in Figure 2 can be represented as \textless{}\textit{groupagent.exe}, \textit{RegOpenKey}, \textit{HKLM\textbackslash System\textbackslash ...\textbackslash SafeBoot\textbackslash Option}\textgreater{}, where \textit{groupagent.exe} opens a registry key (\textless{}\textit{groupagent.exe}, \textit{RegOpenKey}, \textit{HKLM\textbackslash System\textbackslash ...\textbackslash SafeBoot\textbackslash Option}\textgreater{}) through the \textit{RegOpenKey} operation. In this context, \textit{groupagent.exe} and the registry key are system entities, while \textit{RegOpenKey} is the action performed on the registry key.

To process these logs, we use an embedding function that converts the textual log data into numerical vectors that still preserve semantic relationships. SecureBERT~\cite{aghaei2022securebert}, a language model specifically trained on cybersecurity texts,provides deep domain knowledge, making it ideal for embedding system entities and actions into fixed-length vectors. We further apply principal component analysis (PCA)\cite{jolliffe2003principal} to reduce dimensionality. Figure~\ref{fig:triplet_embedding} illustrates the embedding process for an audit log event. The resulting embeddings serve as features of individual events for subsequent tasks such as anomaly detection and technique hunting.

\subsection{Technique Hunting}\label{sec: TTP hunting}

For Technique hunting, SFM incorporates two key functions: managing imbalanced events and detecting Techniques. Given the imbalance between benign and malicious events in real-world scenarios, we first build an anomaly detection model to filter out likely benign events, aiding in the identification of attack patterns. Following this, a neural network model is developed to detect specific attack patterns within the remaining audit events.

\begin{figure}[!htb]
    \centering
    \includegraphics[width=1\linewidth]{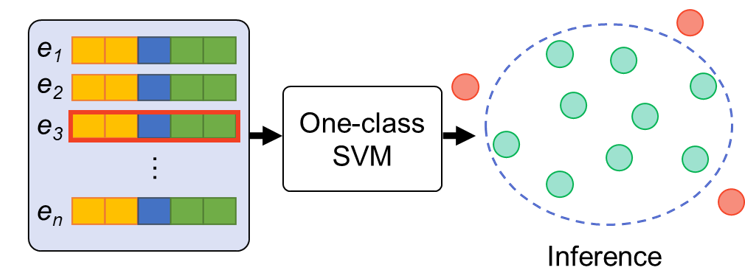}
    \caption{One-class SVM for anomaly detection in event logs.}
    \label{fig:ocsvm}
\end{figure}

\subsubsection{Anomaly Detection}\label{sec: anomaly}
In real-world scenarios, there is often a significant imbalance between attack and benign events. For instance, in the DARPA FiveDirection dataset~\cite{darpa2tc}, the number of attack events is 5,308, compared to 14,005,875 benign events. Machine learning-based classifiers tend to favor the majority class in such imbalanced datasets due to difficulties in learning from the sparse occurrence of the minority class. To mitigate this, we use a one-class support vector machine (SVM)\cite{wenthundersvm18} to preserve likely malicious events. The one-class SVM learns patterns of benign events to establish a decision boundary, where samples deviating from these benign patterns are flagged as potentially malicious, as illustrated in Figure~\ref{fig:ocsvm}.

\begin{figure}[!htb]
    \centering
   \includegraphics[width=0.48\textwidth]{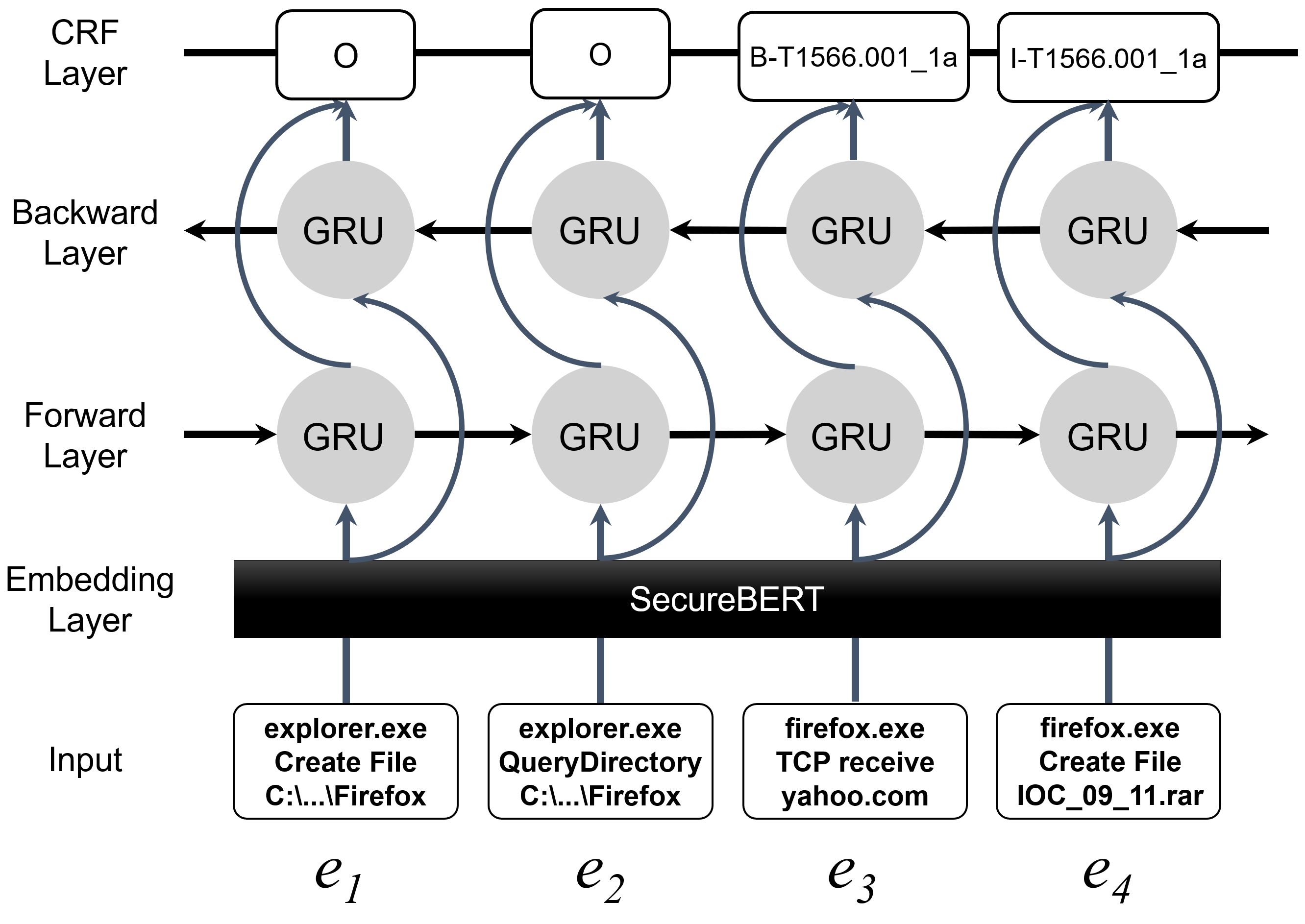}
    \caption{The BiGRU-CRF model for Attack behavior (ability) detection.}
    \label{fig:model}
\end{figure}

\subsubsection{Attack Pattern Detection}
Following anomaly detection, the likely malicious events proceed to attack pattern recognition. Since each Technique may involve multiple abilities, our focus is to identify these abilities within the malicious events. Previous research~\cite{alsaheel2021atlas} has shown that Recurrent Neural Networks (RNNs) are effective for processing attack sequences and distinguishing between attack and non-attack patterns. In this study, we leverage Bidirectional Gated Recurrent Units (BiGRU)~\cite{cho2014learning}, an RNN variant, to identify abilities in a chronological sequence. While a unidirectional GRU processes events in a single direction, bidirectional GRUs enhance this by processing the sequence in both directions, allowing the model to consider the sequence in reverse to improve detection accuracy.

We then incorporate a Conditional Random Field (CRF)~\cite{lafferty2001conditional} to jointly decode labels across the entire sequence. This method enables the collaborative determination of the optimal chain of labels for a given input sequence. Using a CRF layer jointly with BiGRU to model the label sequence, we capture correlations among neighboring labels, resulting in more accurate predictions than independent label decoding.

\subsection{APT Campaign Matching}\label{sec: campaign match}
The discovered ability graph $G_q$ and the campaigns graph $G_c$. are structured such that each node $v$ in $G_q$ represents a detected ability, while a directed edge $e_{i,j}$ indicates a temporal relationship between two abilities involving the same system entities.
The graph $G_q$ constructed at this stage encapsulates the temporal and causal relationships of the discovered attack behaviors, effectively summarizing the attack patterns.
The campaign graph $G_c$ is obtained by retrieving well-documented CTI reports or reputable knowledge bases such as the MITRE ATT\&CK website. In this work, the process is performed manually, but it could potentially be automated in the future.

The problem of matching any subset of $G_q$ with a known campaign graph $G_c$ is framed as a subgraph isomorphism problem, which is NP-complete. This challenge can be tackled using algorithms such as the Ullmann algorithm ~\cite{ullmann1976algorithm} and VF2~\cite{cordella2004sub}.
We observe that nodes within $G_q$ often do not align consistently with nodes in the known campaign $G_c$. This discrepancy suggests that the Technique hunting methods used may result in false positives and false negatives. Consequently, subgraph isomorphism methods may not effectively address the alignment problem.

 To address this challenge, we employ the graph edit distance (GED)~\cite{abu2015exact} to quantify the cost of editing $G_q$ to resemble $G_c$. This method considers the costs associated with deletion, substitution, and insertion operations. The GED is defined as:

\begin{equation}\centering\label{eq:ged}
GED(G_q, G_c) = \min_{o_1, \ldots, o_m \in \gamma(G_q, G_c)} \sum_{i=1}^{m} cost(o_i)
\end{equation}
where $cost(o_i)$ denotes the cost function assessing the strength of an edit operation $o_i$, and $\gamma(G_q, G_c)$ represents the set of editing nodes and paths transforming $G_q$ into $G_c$. 
Since the sizes of campaigns graphs vary, we normalize the $GED$ with Normalized Levenshtein Distance~\cite{gravino2021using} using the following formula. 
\begin{equation}\label{eq:ged_norm}
GED^\prime (G_q, G_c) = \frac{GED(G_q, G_c)}{max(NodeEdge(G_q), NodeEdge(G_c))}
\end{equation}
where NodeEdge() returns the sum of nodes and edges of a given graph. Finally, the normalized edit cost $GED^\prime(G_q, G_c)$ is used as the metric to determine the presence of an APT campaign within $G_q$ and is used for ranking in the empirical study.
\section{Empirical Studies}\label{sec:eval}
\begin{table*}[!htb]
\caption{Overview of APT attack scenarios based on cyber threat intelligence reports.}
\label{tab:known_campaigns}
\centering
\begin{tabular}{|l|l|l|l|l|}
\hline
APT Campaign                            & Attack Stage           & Techniques                                   & Event         & MalEvent    \\ \hline
Higaisa~\cite{Malwarebytes2020Higaisa}  & \{1,2,6,4,4,6,6\}      & PA, MFE, RK, SID, SNCD, MTOS, ST             & 607,416       & 0.005\%     \\ \hline
APT28~\cite{APT28}                      & \{1,2,2,4,4,7\}        & PA, WP, MFE, SID, DLS, EWS                   & 1,203,013     & 1.175\%     \\ \hline
CobaltGroup~\cite{ptsecurity2017Cobalt} & \{1,2,4\}              & PA, RAS, NSD                                 & 961,920       & 0.118\%     \\ \hline
Gamaredon~\cite{CERTEE2021Gamaredon}    & \{1,2,2,6,6,4,4,6,7\}  & PA, WP, MFE, MR, RK, WMI, SID, ST, DF        & 442,729       & 0.013\%     \\ \hline
Patchwork~\cite{Cymmetria2016Patchwork} & \{1,2,3,4,4,4,6,5\}    & PA, PS, BUAC, DLS, UD, SD, RK, RDP           & 155,296       & 9.095\%     \\ \hline

\end{tabular}
    \begin{tablenotes}
        \item PA = phishing Attachment, MFE = Malicious File Execution, RK = Registry Run Keys, SID = System Information Discovery, SNCD = System Network Configuration Discovery, MTOS = Masquerade Task or Service, ST = Scheduled Task, WP = Web Protocols, DLS = Data from Local System, EWS = Exfiltration Over Web Service, RAS = Remote Access Software, NSD = Network Service Discovery, MR = Modify Registry, WMI = Windows Management Instrumentation, DF = Defacement, PS = PowerShell, BUAC = Bypass User Account Control, UD = System Owner/User Discovery, SD = Security Software Discovery, RDP = Remote Desktop Protocol, PEI = Portable Executable Injection, SM = Shortcut Modification, DMT = Disable or Modify Tools, HW = Hidden Window. The subsequent number of a technique represents a distinct ability used to implement that technique~\cite{saga_dataset}.
    \end{tablenotes}
\end{table*}


\subsection{Evaluation Settings}
\noindent
\textbf{Dataset.}
To evaluate SFM's effectiveness in detecting APT attacks, we used a synthetic dataset of APT campaign audit logs~\cite{saga_dataset} and benign data. These logs simulate real-world APT attacks, blending benign user-like actions with malicious events from various APT campaigns, specifically Higaisa, APT28, CobaltGroup, Gamaredon, and Patchwork, based on threat intelligence reports. Malicious events were labeled with the BIO2 scheme, covering 21 attack techniques across 5 campaigns (as shown in Table~\ref{tab:known_campaigns}). This dataset allows for a robust evaluation of detection precision and recall, addressing the scarcity of publicly labeled attack datasets.

\noindent
\textbf{Metrics.}
In line with the event-based methodology for attack pattern detection~\cite{alsaheel2021atlas}, we present the detection results based on events. It is important to note that an event labeled B-XXX or I-XXX \cite{bio22023wiki} is counted as associated with the XXX label. For each attack scenario dataset, we present macro-average precision (P), recall (R), and F1 scores, computed as the arithmetic mean of individual class classification metrics. Note that each scenario involves a distinct set of abilities, and the evaluations consider only the classes within the attack datasets, excluding the label O.

\noindent
\textbf{Implementation details.}
The experiments were conducted on a server equipped with an AMD EPYC 7282 16-core CPU, 1TB of memory, and two NVIDIA A100 GPUs with 80GB of memory each, running on Ubuntu 20.04. The proposed SFM was implemented using Python 3.10.

\begin{table}[!htb]
\caption{Evaluation Results of Threat Hunting}
\label{tab:hunting}
\centering
\setlength{\tabcolsep}{1mm}{
\begin{tabular}{|l|rrr|rrr|}
\hline
             & \multicolumn{3}{c|}{Sigma}                                                            & \multicolumn{3}{c|}{SFM}                                                    \\ \hline
APT Campaign & \multicolumn{1}{c|}{P}       & \multicolumn{1}{c|}{R}       & \multicolumn{1}{c|}{F1} & \multicolumn{1}{c|}{P}       & \multicolumn{1}{c|}{R}       & \multicolumn{1}{c|}{F1} \\ \hline
Higaisa      & \multicolumn{1}{r|}{33.37\%} & \multicolumn{1}{r|}{36.11\%} & 33.40\%                  & \multicolumn{1}{r|}{90.32\%} & \multicolumn{1}{r|}{90.48\%} & 87.00\%                 \\ \hline
APT28        & \multicolumn{1}{r|}{0.00\%}  & \multicolumn{1}{r|}{0.00\%}  & 0.00\%                   & \multicolumn{1}{r|}{56.30\%} & \multicolumn{1}{r|}{62.45\%} & 57.02\%                 \\ \hline
CobaltGroup  & \multicolumn{1}{r|}{0.28\%}  & \multicolumn{1}{r|}{29.75\%} & 0.54\%                   & \multicolumn{1}{r|}{54.82\%} & \multicolumn{1}{r|}{72.31\%} & 58.44\%                 \\ \hline
Gamaredon    & \multicolumn{1}{r|}{25.02\%} & \multicolumn{1}{r|}{17.08\%} & 16.71\%                  & \multicolumn{1}{r|}{73.51\%} & \multicolumn{1}{r|}{77.75\%} & 73.21\%                 \\ \hline
Patchwork    & \multicolumn{1}{r|}{8.13\%}  & \multicolumn{1}{r|}{21.96\%} & 9.14\%                   & \multicolumn{1}{r|}{68.60\%} & \multicolumn{1}{r|}{68.87\%} & 67.55\%                 \\ \hline
Avg.         & \multicolumn{1}{r|}{13.36\%} & \multicolumn{1}{r|}{20.98\%} & 11.96\%                  & \multicolumn{1}{r|}{68.71\%} & \multicolumn{1}{r|}{74.37\%} & 68.64\%                 \\ \hline
\end{tabular}}
\end{table}

\subsection{Evaluation on Effectiveness}

In this section, we compare the performance of the proposed SFM and Sigma~\cite{sigma}, as they are the few available methods for technique hunting. Due to the absence of a standardized benchmark for fine-grained attack pattern detection, five synthetic APT campaigns datasets are used for evaluation. 
Sigma, an open and widely used signature format, facilitates the representation of fine-grained attack patterns. It is supported by a global community of security professionals who collaborate on detection rules. A significant portion of these rules are aligned with the MITRE ATT\&CK framework, and 70 techniques relevant to our attack scenarios are covered by Sigma. Although Sigma may not be a perfect baseline, it provides a valuable point of reference to evaluate the effectiveness of our approach.

The effectiveness of attack pattern detection for each dataset is reported in Table~\ref{tab:hunting}. 
Our methodology exhibits substantial performance improvement compared to Sigma. 
On average, the results demonstrate a precision of 68.71\%, recall of 74.37\%, and an F1 of 68.64\%, while Sigma's results are notably lower at 13.36\%, 20.98\%, and 11.96\%, respectively.
The highest performance was achieved with the Higaisa dataset, achieving 90.32\%, 90.48\% and 87.00\% for precision, recall, and F1, respectively. In contrast, the lowest results were observed in APT28 dataset, with 56.30\%, 62.45\%, and 57.02\% for precision, recall, and F1, respectively.
A significant factor contributing to the high false positive rates is the dramatic imbalance in event classification, as malicious events account for only 1.175\% of the total dtaaset. This disproportion leads to a predominance of benign activities in attack behavior evaluations, causing specific attack signatures to be frequently misidentified. Furthermore, the Sigma rules, while designed by experts, only cover portions of attack behaviors, leaving numerous malicious activities undetected.

\begin{figure}[!htb]
    \centering
    \includegraphics[width=0.8\columnwidth]{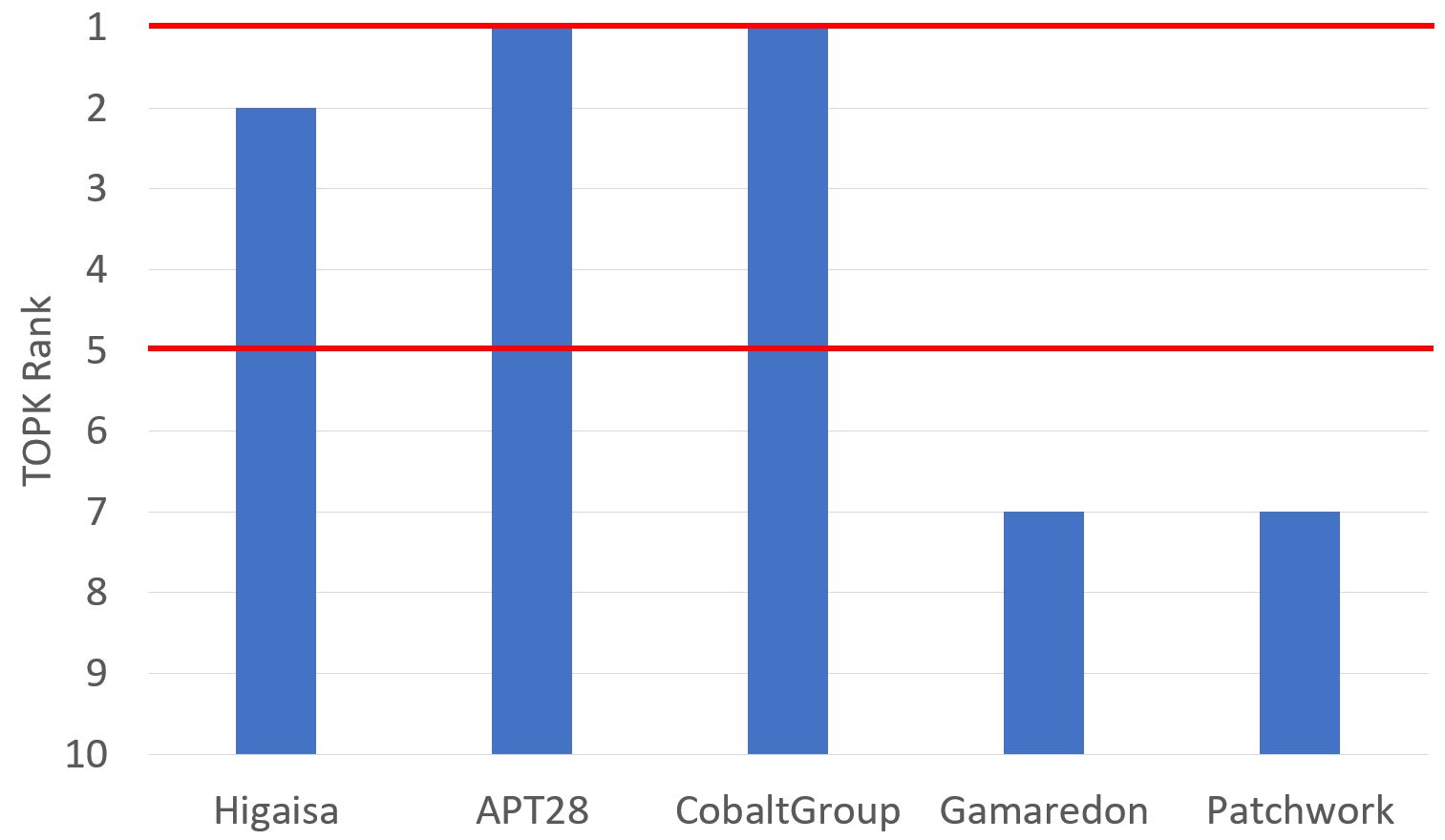}
    \caption{Results on APT campaign matching.}
    \label{fig:TOPK}
    \vspace{-0.6cm}
\end{figure}

\subsection{APT Campaign Detection Evaluation}
The set of known campaigns, denoted as $G_C$, includes five campaigns used to evaluate the effectiveness of campaign matching. As illustrated in Figure 8, the top-k score values for the top 1, 3, and 5 ranked campaigns are 0.4, 0.6, and 0.6, respectively. This suggests that the proposed method effectively establishes detailed temporal causal relationships for attribution, accurately distinguishing the most probable scenarios from the least likely ones. The applied graph edit distance (GED)~\cite{abu2015exact} successfully overlooks minor errors in ability detection. Overall, our approach shows promise in identifying APT campaigns, offering valuable insights for cybersecurity analysts and defenders.

\section{Related work}\label{sec:literature}
\noindent \textbf{Learning-based intrusion detection and analysis.}
Our approach focus on fine-grain technique detection using SecureBERT, a domain-specific pre-trained language model, which enhances contextual understanding and outperform WATSON~\cite{zeng2021watson} and ShadeWatcher~\cite{zengy2022shadewatcher}, which use translation-based embeddings (TransE, TransH) for event structures. Building on methods from Atlas~\cite{alsaheel2021atlas} and DeepAG~\cite{li2022deepag}, We apply recurrent neural networks (RNNs) for audit event processing and incorporate conditional random fields for joint label decoding, which significantly improves detection accuracy.

\noindent
\textbf{Attack pattern recognition.}
Holmes~\cite{milajerdi2019holmes}, RepSheet~\cite{hassan2020tactical}, KRYSTAL~\cite{kurniawan2022krystal} and the Sigma rules search engine~\cite{sigma} are key frameworks in attack pattern detection. Holmes first detects techniques and infers potential attacks; RepSheet reconstructs tactical provenance graphs; KRYSTAL employs RDF-based knowledge graphs to integrate threat detection techniques; and Sigma provides an open signature format for log events. While each contributes to technique detection, these approaches often lack adaptability to evolving threats, underscoring the need for more flexible solutions.

\noindent
\textbf{APT campaign attribution.}
Previous work on APT attribution includes multi-view malware analysis~\cite{sahoo2022cyber}, CSKG4APT~\cite{ren2022cskg4apt}, and APTer~\cite{sachidananda2023apter}. Multi-view analysis attributes APT groups based on malware features like Opcode sequences; CSKG4APT constructs knowledge graphs from real-world APT observations; and APTer maps IDS/IPS alerts to MITRE ATT\&CK techniques for APT attribution. Unlike these approaches, which focus on malware code or CTI reports, we focus on audit events and use graph-matching (graph edit distance) to align techniques with known campaigns, providing an enriched context for interpreting security alerts.

\section{Conclusion}\label{sec:conclusion}
This study presents a machine learning-based method SFM for identifying potential APT threat actors. The evaluation results indicate that SFM successfully detects over 60\% of techniques from system event logs and accurately attributes 86\% of APT campaigns within the top 5 ranks of the known group, highlighting it as a promising approach to APT threat detection and attribution. However, its effectiveness hinges on access to extensive and high-quality campaign knowledge and varied technique implementations. Additionally, system event logs with fine-grained labels are crucial to advance this line of research.

\bibliographystyle{IEEEtran}


\end{document}